\documentclass[a4paper,11pt]{article}

\usepackage[width=15cm,height=23cm]{geometry} 
\usepackage[usenames]{color}
\usepackage{amsfonts,amssymb}
\usepackage{theorem}
\usepackage[dvips]{graphicx}

\long\def\symbolfootnote[#1]#2{\begingroup%
\def\thefootnote{\fnsymbol{footnote}}\footnote[#1]{#2}\endgroup}

\newcommand{\comment}[1]{}

\def\G{\Gamma}

\def\TG{\mathfrak G}

\begin{document}

\vskip 0.4 truecm

\begin{center}{\Large\bf
A Symmetric Approach to the Massive Nonlinear Sigma Model 
\symbolfootnote[2]{\tt This work is supported in part by funds provided by the U.S. Department
of Energy (D.O.E.) under cooperative research agreement \#DE FG02-05ER41360}}
\end{center}
\par
\vskip 1.1 truecm

\large
\rm
\vskip 0.7 truecm
\centerline{ 
R.~Ferrari$^{a,b}$\footnote{e-mail: {\tt ruggero.ferrari@mi.infn.it}}}

\tiny
\medskip
\begin{center}
$^a$
Center for Theoretical Physics\\
Laboratory for Nuclear Science\\
and Department of Physics\\
Massachusetts Institute of Technology\\
Cambridge, Massachusetts 02139
and\\
$^b$
Dip. di Fisica, Universit\`a degli Studi di Milano\\
and INFN, Sez. di Milano\\
via Celoria 16, I-20133 Milano, Italy\\
(MIT-CTP-4167, IFUM-961-FT, July, 2010 )
\end{center}

\normalsize


\begin{quotation}

\rm
 {\Large Abstract:}
In the present paper we extend to the massive case the procedure
of divergences subtraction, previously introduced for the massless
nonlinear sigma model ($D=4$). Perturbative expansion in the number of
loops is successfully constructed. The resulting theory depends
on the Spontaneous Symmetry Breaking parameter $v$, on the mass
$m$ and on the radiative correction parameter $\Lambda$. Fermions are not
considered in the present work. $SU(2)\otimes SU(2)$ is the group
used. 
\end{quotation}

\newpage
\section{Introduction}
\label{sec:intro}
It is of paramount importance to establish a symmetric subtraction strategy
for the divergences of
the \underline{massive} nonlinear sigma model. In fact this model provides a  viable
infrared regulator, a useful phenomenological theory
and finally a template to approach more complex theories, as
the nonabelian gauge theories in the 't Hooft gauge.
Moreover a  simulation on a lattice is free from the artifacts related
to the presence of the zero modes. 
\par
The subtraction strategy for the ultraviolet divergences,
devised for the massless nonlinear sigma model \cite{Ferrari:2005ii},
can be successfully extended to the model with non zero mass, although  chirality is broken.
\par 
A powerful  Local Functional
Equation (LFE) for the generating functionals (of the Green- 
and one-particle-irreducible-functions (1-PI)) ensues from
the invariance properties of the
path integral measure. By using the LFE, one
can derive the complete hierarchy relations among the ancestor amplitudes
(i.e. with no pion fields) and the descendant amplitudes (i.e.
with at least one pion field) and get the full control of the
divergences by a finite number of divergent amplitudes
at each order in the loop expansion. The counterterms are chosen according to the
symmetry properties of the effective action (not of the action!) fixed by the LFE. 
Pure pole subtraction in dimensional
regularization turns out to be the right thing to do, in order
to satisfy the LFE.
\par
The  physical parameters of the model are those of
the classical action augmented by the scale of the radiative
corrections $\Lambda$ (introduced later on).
\par
To compare with the Chiral Perturbation Theory (ChPT) 
\cite{Weinberg:1978kz}-\cite{Bijnens:2009zi}, we stress
that the subtracted amplitudes depend on
a fixed number of parameters. Moreover another important
feature characterizes the present approach: $\Lambda$ is a physical parameter,
while in ChPT
the requirement of independence from 
$\Lambda$ leads to a Renormalization Group Equation.
\section{The Model}
\label{sec:model}
The generating functionals $Z$ and $W$ are introduced via the path integral
\begin{eqnarray}&&
Z[\vec K,K_0,\vec J_\mu,N_{AB}] = \exp iW
\simeq \int {\cal D}[\phi]\frac{1}{2\phi_0}
\exp\biggl[i S 
\nonumber\\&&
+ \Lambda^{D-4} \int d^D x \Bigl(\phi_0 K_0 +\phi_A N_{AB}\phi_B\Bigr) + \int d^D x\phi_a K_a
\biggr].
\label{model.1}
\end{eqnarray}
The classical action is
\begin{eqnarray}&&
 S=\Lambda^{D-4} \int d^D x \frac{v^2}{2}\Bigl[\frac{1}{2}~Tr~\Bigl\{ (F_{\mu}-J_{\mu})^2 \Bigr\}
 - m^2 \vec\phi^2 \Bigr]
\nonumber\\&& 
= \Lambda^{D-4} \int d^D x \frac{v^2}{2}\Bigl[\frac{1}{4}\Bigl(F_{a\mu}-J_{a\mu}\Bigr)^2
 - m^2 \vec\phi^2 \Bigr]
\nonumber\\&&
= \Lambda^{D-4} \int d^D x \frac{v^2}{2}\Bigl(\partial_\mu\phi_A\partial^\mu\phi_A -
\frac{1}{2} F_{a\mu}J_{a}^\mu+ \frac{1}{4} J^2 - m^2 \vec\phi^2\Bigr). 
\label{model.2}
\end{eqnarray}
where
\begin{eqnarray}&&
\phi_0= \sqrt{1-\vec\phi\,^2}
\nonumber\\&&
\Omega = \phi_A\tau_A=\phi_0 +i \phi_a\tau_a , \quad \Omega \in SU(2)
\nonumber\\&&
F_\mu = \frac{\tau_a}{2}F_{a\mu}\equiv i \Omega\partial_\mu\Omega^\dagger
\nonumber\\&&
F_{a\mu} = 2(\phi_0\partial_\mu\phi_a
-\partial_\mu\phi_0\phi_a + \epsilon_{abc}\partial_\mu\phi_b \phi_c)
\nonumber\\&&
\tau_A\equiv \{1, i \vec\tau\}.
\label{model.3}
\end{eqnarray}
Capital letter indexes run over $\{0,1,2,3\}$ while lower cases over $\{1,2,3\}$. 
The fact that
the fields $\phi_0, \vec\phi$ have null canonical dimension is only a matter of choice.
$N_{AB}$ are  real independent sources introduced in order to account for the
extra composite operators generated by the local chiral transformations. By construction
we have
\begin{eqnarray}
Z[\vec K,K_0,\vec J_\mu,N_{AB}] = 
Z[\vec K,K_0,\vec J_\mu,N_{BA}],
\label{model.1.1}
\end{eqnarray}
i.e. $Z$ does not depend on the antisymmetric part of the
matrix $N_{BA}$. 
\section{The Local Functional Equation}
\label{sec:lfe}
The measure in the path integral (\ref{model.1}) is invariant under the
local left-transformation
\begin{eqnarray}
\Omega'=U(\vec \omega)\Omega
\label{lfe.1}
\end{eqnarray}
i.e.
\begin{eqnarray}&&
\delta \phi_0 = -  \frac{\omega_a(x)}{2}\phi_a
\nonumber\\&&
\delta \phi_a = \frac{ \omega_a(x)}{2}\phi_0 + \frac{
\omega_c(x)}{2}
\epsilon_{abc}\phi_b
\label{lfe.2}
\end{eqnarray}
for infinitesimal parameters $\vec\omega$.
\par\noindent
The path integral is invariant under the field-coordinate
transformation (\ref{lfe.2}), thus we get an identity for  the
connected amplitude functional 
\begin{eqnarray}&&
\int d^Dx\Biggl\langle
\Lambda^{D-4} \Biggl\{v^2\Bigl[
\frac{1}{4}(F_{a\mu}- J_{a\mu})(\epsilon_{abc}F_b^\mu\omega_c+\partial^\mu \omega_a )
-m^2 \frac{\omega_a}{2}\phi_a \phi_0 \Bigr]
\nonumber\\&&
-\frac{\omega_a}{2}\phi_a K_0 - \frac{\omega_a}{2}\phi_a  N_{0B}\phi_B 
+ \frac{\omega_a}{2} \phi_0 N_{aB}\phi_B
+\frac{\omega_c}{2}\epsilon_{abc}\phi_b  N_{aB}\phi_B
\nonumber\\&&
 -  \phi_A  N_{A0}\frac{\omega_b}{2}\phi_b
+ \phi_A  N_{Aa}\frac{\omega_a}{2}\phi_0
+ \phi_A  N_{Ab}\frac{\omega_c}{2}\epsilon_{bb'c}\phi_{b'}
\Biggr\}
\nonumber\\&&
+\frac{\omega_a}{2} \phi_0 K_a  +\frac{\omega_c}{2}\epsilon_{abc}\phi_b K_a
\Biggr\rangle_C  =0.
\label{lfe.3}
\end{eqnarray}
The brackets $\langle  \cdot \rangle$ denote the mean value over the paths according to
eq. (\ref{model.1}).
Then by using the symmetry (\ref{model.1.1})
\begin{eqnarray}&&
\Biggl\{
(-\epsilon_{abc}J_c^\mu+\delta_{ab}\partial^\mu  )\frac{\delta}{\delta J_b^\mu}
-m^2 \frac{v^2}{2}\frac{\delta}{\delta N_{a0}}
\nonumber\\&&
- \frac{1}{2} N_{0B}\frac{\delta}{\delta N_{aB}} 
+  \frac{1}{2} N_{aB}\frac{\delta}{\delta N_{0B}}
- \frac{1}{2}\epsilon_{abc} N_{cB}\frac{\delta}{\delta N_{bB}}
\nonumber\\&&
- \frac{1}{2} N_{A0}\frac{\delta}{\delta N_{Aa}} 
+  \frac{1}{2} N_{Aa}\frac{\delta}{\delta N_{A0}}
- \frac{1}{2}\epsilon_{abc} N_{Ac}\frac{\delta}{\delta N_{Ab}}
\nonumber\\&&
-\frac{\Lambda^{D-4} }{2} K_0 \frac{\delta}{\delta K_a}
+\frac{1}{2\Lambda^{D-4} } K_a \frac{\delta}{\delta K_0} +\frac{1}{2}\epsilon_{abc} K_b\frac{\delta}{\delta K_c}
\Biggr\} \, W  =0.
\label{lfe.4}
\end{eqnarray}
We introduce the generators $ L_{cAB}$ of the transformations (\ref{lfe.2})
\begin{eqnarray}&&
\delta\phi_A = \frac{1}{2}\omega_c L_{cAB} \phi_B,\quad
L_{c0b}= -\delta_{cb}, L_{ca0}= \delta_{ca}, L_{cab}=\epsilon_{abc}
\nonumber\\&&
\omega_c L_c= 
\left(
\begin{array}{llll}
0&-\omega_1 & - \omega_2 & - \omega_3 \\
\omega_1&0&\omega_3&-\omega_2\\
\omega_2&-\omega_3&0&\omega_1\\
\omega_3&\omega_2&-\omega_1&0\\
\end{array}
\right)
\nonumber\\&&
[L_a,L_b] = - 2 \epsilon_{abc}L_c
\label{lfe.4.1}
\end{eqnarray}
and we rewrite eq. (\ref{lfe.4}) in the compact form
\begin{eqnarray}&&
\Biggl\{
(-\epsilon_{abc}J_c^\mu+\delta_{ab}\partial^\mu  )\frac{\delta}{\delta J_b^\mu}
-m^2 \frac{v^2}{2}\frac{\delta}{\delta N_{a0}}
\nonumber\\&&
- \frac{1}{2} \Bigl( L_{aAA'}N_{A'B}+L_{aBB'} N_{AB'}\Bigr)\frac{\delta}{\delta N_{AB}} 
\nonumber\\&&
-\frac{\Lambda^{D-4} }{2} K_0 \frac{\delta}{\delta K_a}
+\frac{1}{2\Lambda^{D-4} } K_a \frac{\delta}{\delta K_0} +\frac{1}{2}\epsilon_{abc} K_b\frac{\delta}{\delta K_c}
\Biggr\} \, W  =0.
\label{lfe.4.2}
\end{eqnarray}
For the effective action (1-PI generating functional) one gets
\begin{eqnarray}&&
-\partial^\mu \Gamma_{J_a^\mu} - \epsilon_{abc}J_b^\mu \Gamma_{J_c^\mu} 
+ \frac{v^2}{2} m^2 \Gamma_{N_{0a}}
+ \frac{1}{2} \Bigl( L_{aAA'}N_{A'B}+L_{aBB'} N_{AB'}\Bigr)\Gamma_{ N_{AB}}
\nonumber\\&&
+ \frac{\Lambda^{D-4} }{2} K_0 \phi_a 
+\frac{1}{2\Lambda^{D-4} }  \Gamma_{ K_0} \Gamma_{\phi_a}
- 
\frac{1}{2}\epsilon_{abc}\phi_b \Gamma_{\phi_c}
 \,   =0,
\label{lfe.5}
\end{eqnarray}
where we use the notation
\begin{eqnarray}
\Gamma_X= 
 \frac{\delta }{\delta X} \Gamma .
\label{lfe.5.1}
\end{eqnarray}
%

%
\section{The Subtraction Strategy at $D=4$}
\label{sec:lin}
The LFE (\ref{lfe.5}) is the tool we use in order
to make finite the massive nonlinear sigma model. 
We work in the framework of the loop expansion of $\Gamma$ in eq. (\ref{lfe.5}). 
\par
This strategy has been employed with success for the massless nonlinear sigma model
\cite{Ferrari:2005va}-\cite{Bettinelli:2007kc}, for the massive Yang-Mills in the Landau gauge
\cite{Bettinelli:2007tq},\cite{Bettinelli:2007cy}, for a Higgsless Electroweak model also
in the Landau gauge  \cite{Bettinelli:2008ey}-\cite{Bettinelli:2009wu}. 
Therefore in models where dimensional regularization allows
to drop the tadpoles involving the scalars fields $\vec\phi$. In the massive nonlinear sigma model
the tadpoles play an important r\^ole and therefore it is necessary to extend the
formalism to this case. 
\par
At the tree level  $\Gamma^{(0)}$ is a solution of eq. (\ref{lfe.5}) by construction.
Dimensional regularization yields radiative corrections which do not generate any
anomaly. The proof of this property is sketched in Ref. \cite{Ferrari:2005ii}
and displayed in Ref. \cite{Bettinelli:2007zn}.
The Feynman rules are provided
by the classical action $\Gamma^{(0)}$, while the counterterms $\hat\Gamma^{(k)}$
are introduced via the effective action which must obey the LFE (\ref{lfe.5}).
Therefore what matters are the symmetry properties of $\Gamma$ and not those
of the action. 
\par
If the procedure
of subtraction of infinities by means of the counterterms $\hat\Gamma^{(k)}$
has been carried out successfully up to order $n-1$, then
\begin{eqnarray}&& \!\!\!\!\!\!\!\!\!\!\!\!
-\partial^\mu \Gamma^{(k)}_{J_a^\mu} - \epsilon_{abc}J_b^\mu \Gamma^{(k)}_{J_c^\mu} 
+ \frac{v^2}{2} m^2 \Gamma^{(k)}_{N_{0a}}
+ \frac{1}{2} \Bigl( L_{aAA'}N_{A'B}+L_{aBB'} N_{AB'}\Bigr)\Gamma^{(k)}_{ N_{AB}}
\nonumber\\&&
\!\!\!\!\!\!\!\!\!\!\!\!
+\frac{1}{2\Lambda^{D-4} }  \Gamma^{(0)}_{\phi_a} \Gamma^{(k)}_{ K_0} 
+\frac{1}{2} \phi_0 \Gamma^{(k)}_{\phi_a}
- 
\frac{1}{2}\epsilon_{abc}\phi_b \Gamma^{(k)}_{\phi_c}
 \,    +\sum_{j=1}^{k-1}
\frac{1}{2\Lambda^{D-4} }  \Gamma^{(k-j)}_{ K_0} \Gamma^{(j)}_{\phi_a} = 0,\,\, \forall k<n.
\label{lin.1}
\end{eqnarray}
At order $n$ we expect a violation of eq. (\ref{lfe.5})
\begin{eqnarray}&&
-\partial^\mu \Gamma^{(n)}_{J_a^\mu} - \epsilon_{abc}J_b^\mu \Gamma^{(n)}_{J_c^\mu} 
+ \frac{v^2}{2} m^2 \Gamma^{(n)}_{N_{0a}}
+ \frac{1}{2} \Bigl( L_{aAA'}N_{A'B}+L_{aBB'} N_{AB'}\Bigr)\Gamma^{(k)}_{ N_{AB}}
\nonumber\\&&
+\frac{1}{2\Lambda^{D-4} }  \Gamma^{(0)}_{\phi_a} \Gamma^{(n)}_{ K_0} 
+\frac{1}{2} \phi_0 \Gamma^{(n)}_{\phi_a}
- 
\frac{1}{2}\epsilon_{abc}\phi_b \Gamma^{(n)}_{\phi_c}
 \,    +\sum_{j=1}^{n-1}
\frac{1}{2\Lambda^{D-4} }  \Gamma^{(n-j)}_{ K_0} \Gamma^{(j)}_{\phi_a}
\nonumber\\&&
 = 
\sum_{j=1}^{n-1}
\frac{1}{2\Lambda^{D-4}} \widehat{ \Gamma}^{(n-j)}_{ K_0} \widehat{ \Gamma}^{(j)}_{\phi_a}.
\label{lin.1.1}
\end{eqnarray}
By assumption $\Gamma^{(k<n)}$ is finite, 
thus the removal of the poles in the Laurent expansion of
\begin{eqnarray}
\frac{1}{\Lambda^{D-4} } \Gamma^{(n)}
\label{lin.1.1.1}
\end{eqnarray}
is a strategy that maintains the validity of the LFE,
since the RHS of eq. (\ref{lin.1.1}) is a pure pole part, when the
normalization (\ref{lin.1.1.1}) is used.
The finite part in the limit
$D=4$ is the subtracted  amplitude. It should be stressed that no further
finite subtraction is allowed (even those keeping eq. (\ref{lin.1.1}) unchanged). In fact
by adding extra counterterms one modifies the pure pole structure of the breaking terms.
For instance on-shell renormalization is not a doable procedure.
Finally the spotted  counterterms $\widehat{ \Gamma}^{(n)}$ 
obey the equation
\begin{eqnarray}&&
-\partial^\mu \widehat{ \Gamma}^{(n)}_{J_a^\mu} - \epsilon_{abc}J_b^\mu \widehat{ \Gamma}^{(n)}_{J_c^\mu} 
+ \frac{v^2}{2} m^2 \widehat{ \Gamma}^{(n)}_{N_{0a}}
+ \frac{1}{2} \Bigl( L_{aAA'}N_{A'B}+L_{aBB'} N_{AB'}\Bigr)\widehat{\Gamma}^{(k)}_{ N_{AB}}
\nonumber\\&&
+\frac{1}{2\Lambda^{D-4} }  \widehat{ \Gamma}^{(0)}_{\phi_a} \widehat{ \Gamma}^{(n)}_{ K_0} 
+\frac{1}{2} \phi_0 \widehat{ \Gamma}^{(n)}_{\phi_a}
- 
\frac{1}{2}\epsilon_{abc}\phi_b \widehat{ \Gamma}^{(n)}_{\phi_c}
 +
 \sum_{j=1}^{n-1}
\frac{1}{2\Lambda^{D-4} } \widehat{ \Gamma}^{(n-j)}_{ K_0} \widehat{ \Gamma}^{(j)}_{\phi_a}=0.
\label{lin.1.2}
\end{eqnarray}
The last equation (\ref{lin.1.2}) provides a {\sl a posteriori} explanation of why the
breaking term takes the form exhibited in (\ref{lin.1.1}). 
A direct proof is provided in Ref. \cite{Bettinelli:2007zn}. The subtraction strategy has been
tested for a solvable model in Ref. \cite{Ferrari:2009uj}.
%
%
\section{The Algebraic Aspects of the LFE}
\label{sec:alg}
It is convenient to define a new functional
\begin{eqnarray}
\TG[\phi_a, J_a^\mu,K_0,N_{00},N_{0a}, N_{ab}]
\equiv \Gamma[\phi_a, J_a^\mu,K_0,N_{00}{-}\frac{1}{2}v^2m^2,N_{0a}, N_{ab}]
\label{alg.2}
\end{eqnarray}
in presence of the \underline{nontrivial background}
\begin{eqnarray}&&
\phi_a = J_a^\mu = K_0 = N_{0a}= N_{ab}=0
\nonumber\\&&
N_{00} =\frac{1}{2} m^2 v^2
\label{alg.2.1}
\end{eqnarray}
and of the boundary conditions
\begin{eqnarray}&&
\frac{\delta \TG^{(0)}}{\delta K_0}\biggl |_{\phi_a = J_a^\mu = K_0 = N_{0a}= N_{ab}=[N_{00} {-}\frac{1}{2}v^2 m^2]=0} = \Lambda^{D-4}
\nonumber\\&&
\frac{\delta \TG^{(0)}}{\delta N_{00}}\biggl |_{\phi_a = J_a^\mu = K_0 = N_{0a}= N_{ab}=[N_{00} {-}\frac{1}{2}v^2 m^2]=0} = \Lambda^{D-4}.
\label{alg.2.1.bou}
\end{eqnarray}
By using $ \TG$ we get rid of the mass term in eq. (\ref{lfe.5}) and all the subsequent equations. 
For the counterterms $\widehat\TG^{(n)}$ eq. (\ref{lin.1.2}) gives
\begin{eqnarray}
{\cal S}_a\widehat\TG^{(n)}
 \,   = - \sum_{j=1}^{n-1}
\frac{1}{2\Lambda^{D-4} } \widehat \TG^{(n-j)}_{ K_0} \widehat\TG^{(j)}_{\phi_a},
\label{alg.3}
\end{eqnarray}
where
\begin{eqnarray}&&
{\cal S}_a \equiv
-\partial^\mu \frac{\delta}{\delta J_a^\mu} - \epsilon_{abc}J_b^\mu 
\frac{\delta}{\delta J_c^\mu}
+ \frac{1}{2} \Bigl( L_{aAA'}N_{A'B}+L_{aBB'} N_{AB'}\Bigr)\frac{\delta}{\delta N_{AB}} 
\nonumber\\&&
+\frac{1}{2\Lambda^{D-4}}  \TG^{(0)}_{\phi_a} \frac{\delta}{K_0}
+\frac{1}{2} \phi_0 \frac{\delta}{\phi_a}
- 
\frac{1}{2}\epsilon_{abc}\phi_b \frac{\delta}{\phi_c}.
\label{alg.4}
\end{eqnarray}
It is convenient to introduce the operators
\begin{eqnarray}&&
{\cal S}^J_a \equiv
-\partial^\mu \frac{\delta}{\delta J_a^\mu} - \epsilon_{abc}J_b^\mu 
\frac{\delta}{\delta J_c^\mu}
\nonumber\\&&
{\cal S}^N_a \equiv
\frac{1}{2} \Bigl( L_{aAC}N_{CB}+ L_{aBD}N_{AD}\Bigr)\frac{\delta}{\delta N_{AB}} 
\nonumber\\&&
{\cal S}^\phi_a \equiv
\frac{1}{2} \phi_0 \frac{\delta}{\delta\phi_a}
-
\frac{1}{2}\epsilon_{abc}\phi_b \frac{\delta}{\delta\phi_c}
\nonumber\\&&
{\cal S}^\TG_a \equiv
\frac{1}{2\Lambda^{D-4}} \TG^{(0)}_{\phi_a}
\frac{\delta}{\delta K_0}.
\label{alg.4.1}
\end{eqnarray}
It is straightforward to verify the following relations
\begin{eqnarray}&&
[{\cal S}^J_a(x) , {\cal S}^J_{a'}(y)]
=   \epsilon_{a a'c}\delta(x-y){\cal S}^J_c(x)
\nonumber\\&&
[{\cal S}^N_a(x) , {\cal S}^N_{a'}(y)]
=   \epsilon_{a a'c}\delta(x-y){\cal S}^N_c(x)
\nonumber\\&&
[{\cal S}^\phi_a(x) , {\cal S}^\phi_{a'}(y)]
=   \epsilon_{a a'c}\delta(x-y){\cal S}^\phi_c(x)
\label{alg.10.1.1}
\end{eqnarray}
and finally
\begin{eqnarray}
[{\cal S}_a(x) , {\cal S}_{a'}(y)]
=   \epsilon_{a a'c}\delta(x-y){\cal S}_c(x).
\label{alg.10.1}
\end{eqnarray}
%
\subsection{The Consistency Condition}
From eqs. (\ref{alg.3}) and  (\ref{alg.10.1}) we get the consistency condition
that has to be obeyed by the counterterms
\begin{eqnarray}&&
\Delta_a(x) \equiv  - \sum_{j=1}^{n-1}\,
 \widehat \TG^{(n-j)}_{ K_0(x)}\,  \widehat\TG^{(j)}_{\phi_a(x)}
\nonumber\\&&
{\cal S}_a(x) \Delta_{a'}(y)
 -
{\cal S}_{a'}(y) \Delta_a(x)
=\delta(x-y) \epsilon_{aa'c}~\Delta_c(x).
\label{alg.8.1}
\end{eqnarray}
%
\subsection{The Local  Solutions: the bleaching Method}
The counterterms $\widehat \TG^{(n)}$ are given by linear combinations of local monomials
constructed in terms of fields $\vec\phi$ and sources $K_0, J_a^\mu, N_{AB}$.
The construction of the counterterms proceeds via the evaluation
of the pole parts of the amplitudes as  in eq. (\ref{lin.1.1.1})
and consequently the fixing of the coefficients of the general solution
of eq. (\ref{alg.3}). For one-loop corrections one needs just the solution
for the associated homogeneous equation.
\par
This task is made easy if we replace the above mentioned variables by
suitably chosen composite local invariants. The general procedure 
in Ref. \cite{Bettinelli:2007kc} is here applied straightforwardly.
\par
The local invariant solutions can be constructed with the following fields
and sources
\begin{eqnarray}&&
F_a^\mu- J_a^\mu
\nonumber\\&&
D^\mu[F] (F-J)^\nu|_{ab} =(\partial_\mu \delta_{ab}-\epsilon_{abc}F_c^\mu) (F_b^\nu- J_b^\nu)
=\partial_\mu \delta_{ab}-\epsilon_{abc}J_c^\mu (F_b^\nu- J_b^\nu)
\nonumber\\&&
\phi_A \equiv\{\phi_0,\phi_a\}
\nonumber\\&&
K_A \equiv \{ K_0,-\frac{1}{\Lambda^{D-4} }\TG^{(0)}_{\phi_a}\},
\qquad -\frac{1}{\Lambda^{D-4} }\TG^{(0)}_{\phi_a}\Big|_{\vec\phi=0}
= - \frac{v^2}{2}\partial_\mu J_a^\mu - N_{0a} - N_{a0}
\nonumber\\&&
N_{AB}.
\label{alg.5}
\end{eqnarray}
(Notice $D^\mu[F] (F-J)^\nu|_{ab}-D^\mu[J] (F-J)^\nu|_{ab}
=-\epsilon_{abc}(F_c^\mu- J_c^\mu)(F_b^\nu- J_b^\nu)$).
\par\noindent
$K_A$ and $\phi_A$ transform in the same way. In fact
\begin{eqnarray}&&
{\cal S}_a (x)\TG^{(0)}_{\phi_b(y)}
 =
\frac{1}{2}\delta(x-y) \Bigl\{-\delta_{ab}\Lambda^{D-4}  K_0 (y)
+\epsilon_{bca} \TG^{(0)}_c(y)  
\Bigr \}
\nonumber\\&&
{\cal S}_a (x)K_0(y)=\delta(x-y)\frac{1}{2\Lambda^{D-4} }
\TG^{(0)}_{\phi_a(y)}.
\label{alg.5.1}
\end{eqnarray}
The bleached  variables can be constructed 
according to the transformation
properties of the fields  in eq. (\ref{alg.5}). For instance
{
\begin{eqnarray}&&
{\mathfrak J}_{\mu} \equiv \Omega^\dagger (J_{\mu}-F_{\mu})\Omega
=\Omega^\dagger J_{\mu}\Omega + i \Omega^\dagger\partial_\mu\Omega\, , 
\nonumber\\&&
\partial_\mu{\mathfrak J}_{\nu} =\Omega^\dagger \left(\partial_\mu+
\Omega\partial_\mu\Omega^\dagger 
\right)(J_{\nu}-F_{\nu})\Omega=\Omega^\dagger{\cal D}_\mu[F](J_{\nu}-F_{\nu})\Omega\, , 
\nonumber\\&&
{\mathfrak K}_0 \equiv K_A \phi_A =
K_0\phi_0 - \frac{1}{\Lambda^{D-4} }\TG^{(0)}_{\phi_a} \phi_a\, , 
\nonumber\\&&
{\mathfrak N}_{\alpha \beta, \rho \sigma}= \Omega^\dagger_{\alpha\alpha'}(\tau_A)_{\alpha' \rho}
(\tau_B^\dagger)_{\beta\beta' }\Omega_{\beta'\sigma} N_{AB}.
\label{alg.6}
\end{eqnarray}
}
The bleaching yields ${\cal S}_a$-invariant local variables. Moreover the mapping is
invertible. 
\par
The actual construction of the counterterms can profit of further properties of
the LFE that limit the number of independent divergent amplitudes.  The next Section
deals with this feature.

%
\section{Hierarchy and Weak Power Counting}
\label{sec:h}
Eq. (\ref{lfe.5}) is nonlinear in $\Gamma$. This allows us to grade
the 1PI functions in a hierarchical way according to the
number of external $\phi$ - legs. In fact we have
\begin{eqnarray}&&
\Gamma_{\phi_a} = \frac{\Lambda^{D-4} } { \Gamma_{ K_0} }
\Biggl\{
\epsilon_{abc}\phi_b \Gamma_{\phi_c}
+2\partial^\mu \Gamma_{J_a^\mu} + 2\epsilon_{abc}J_b^\mu \Gamma_{J_c^\mu} 
\nonumber\\&&
-v^2 m^2 \Gamma_{N_{0a}}
- \Lambda^{D-4} K_0 \phi_a 
-\Bigl( L_{aAA'}N_{A'B}+L_{aBB'} N_{AB'}\Bigr)\Gamma_{ N_{AB}}
\Biggr\}
\label{h.1}
\end{eqnarray}
The derivative of eq.(\ref{h.1}) with respect to any
ancestor variable ($J_a^\mu, K_0, N_{AB}  $) yields all
the descendant amplitudes involving one $\phi-$ field.
In a recursive way one obtains all the descendant
amplitudes from the ancestor ones (hierarchy). 
\par
This result is
very important because, at fixed order in the loop
expansion, the number of independent divergent ancestor amplitudes
is finite. In fact,
by simple dimensional analysis, one can show that the superficial degree
of divergence of a 1-PI graph $G$ for ancestor and descendant variables 
is bounded by
\begin{eqnarray}
\delta(G)\leq n_L(D-2)+2 -  N_J-2(N_K+N_N),
\label{h.2}
\end{eqnarray}
where $n_L$ is the number of loops and $N_J,N_K,N_N$
are the numbers of insertions of the ancestor variables
$J_a^\mu, K_0, N_{AB}  $. Thus for fix $n_L$ the number
of independent divergent ancestor amplitudes is finite.
The bound in eq. (\ref{h.2}) does not depend on the number of external
$\phi$ - legs; therefore, if the ancestor amplitudes are
divergent, an infinite number of descendant will also be divergent.
The divergent parts of the descendant amplitudes will not be independent, 
due to the hierarchy property.
\par\noindent
The bound (\ref{h.2}) comes from the following two relations
\begin{eqnarray}&&
\delta(G)= n_L D -2 I  +  N_J + \sum_k k V_k
\nonumber\\&&
n_L = I - N_J - (N_K+N_N) -\sum_k  V_k + 1
\label{h.2.7}
\end{eqnarray}
where $V_k$ is the number of vertexes with $k$ derivatives
and $I$ the number of internal lines.
The inequality in eq. (\ref{h.2}) comes from the fact that for the
unsubtracted theory $k\le 2$. 
\par
The WPC criterion consists in building the classical
action $\Gamma^{(0)}$ such that the bound (\ref{h.2}) is obeyed.
It is not necessary to introduce in $\Gamma^{(0)}$ all possible terms
that are allowed by the WPC criterion. For instance it is not necessary
to introduce a $\phi^4$ interaction in the model considered in
the present paper. This approach is at variance with the algebraic
renormalization procedure where the Power Counting theorem requires
that all allowed couplings should enter with independent parameters. 
\par
The WPC theorem says that the bound (\ref{h.2})
is stable under the subtraction procedure described 
in Section \ref{sec:lin}. The proof goes as follows.  
The counterterm  $\widehat\Gamma^{(k)}$,
of order $k$ in the loop expansion,
is a finite sum of local monomials built with  $J_a^\mu, K_0, N_{AB}  $
sources, space-time derivatives and $N_\phi$  $\phi$ - fields. 
Each monomial ${\cal M}^{(k)}(J_a^\mu, K_0, N_{AB},N_\phi,n_d)$
entering in   $\widehat\Gamma^{(k)}$ is constructed from graphs that
obey eq. (\ref{h.2}) and therefore it satisfies the condition
\begin{eqnarray}
 k (D-2)+2  - n_J- 2(n_K+n_N)- n_d \ge 0 ,
\label{h.3}
\end{eqnarray}
where $n_d$ is the number of derivatives and $n_J,n_K,n_N$ are
the numbers of times the sources enter in the monomial.
When one of this monomial is inserted in a graph, then the
relations in eqs. (\ref{h.2.7}) are modified by an extra vertex
with $n_d$ derivatives. Moreover the numbers of external
sources become $N_J-n_J, N_K-n_K, N_N-n_N$. The superficial
degree of divergence becomes
\begin{eqnarray}&&
\delta(G)\leq (n_L-k)(D-2)  + n_d -  (N_J-n_J)-2[(N_K-n_K)+(N_N-n_N)] 
\nonumber\\&&
\leq n_L(D-2)+2 -  N_J-2(N_K+N_N)
\label{h.4}
\end{eqnarray}
by using eq. (\ref{h.3}) in the last step.
Thus WPC is stable under subtraction of divergences.
%
\section{Hierarchical Relations and Feynman Rules}
\label{sec:h2}
%
In this Section we discuss some of the hierarchical relations
for the two-point functions. This is an example. More
relations and Feynman rules are given in Appendix \ref{app:A}. 
\par
By successive differentiation of eq. (\ref{lfe.5}) we get
\begin{eqnarray}&&
\partial^\mu \Gamma_{J_a^\mu J_b^\nu} 
- \frac{v^2}{2}m^2 \Gamma_{N_{0a}J_b^\nu}
-\frac{1}{2\Lambda^{D-4} }  \Gamma_{ K_0} \Gamma_{\phi_aJ_b^\nu}
 \,   =0.
 \nonumber\\&&
\partial^\mu \Gamma_{J_a^\mu N_{0b}} 
- \frac{v^2}{2}m^2 \Gamma_{N_{0a} N_{0b}}
{-\frac{1}{2} \Gamma_{ N_{ab}}}
+ \frac{1}{2} \delta_{ab}\Gamma_{ N_{00}}
+ \frac{1}{2}\epsilon_{abc} \Gamma_{ N_{0c}}
 \nonumber\\&&
-\frac{1}{2\Lambda^{D-4} }  \Gamma_{ K_0} \Gamma_{\phi_a  N_{0b}}
 \,   =0,
\label{h2.1}
\end{eqnarray}
%
%
\begin{eqnarray}
\partial^\mu \Gamma_{J_a^\mu \phi_b}
- \frac{v^2}{2}m^2 \Gamma_{N_{0a} \phi_b}
-\frac{1}{2\Lambda^{D-4} }  \Gamma_{ K_0} \Gamma_{\phi_a  \phi_b}
 \,   =0
\label{h2.2}
\end{eqnarray}
and finally ($\partial^\mu \Gamma_{ K_0}=0$)
\begin{eqnarray}&&
\!\!\!\!\!\!\!\!\!\!\!\!
\Gamma_{\phi_a  \phi_b} = \frac{2\Lambda^{D-4} }{\Gamma_{ K_0} }
\Biggl(
\partial^\mu \Gamma_{J_a^\mu \phi_b}
- \frac{v^2}{2}m^2 \Gamma_{N_{0a} \phi_b}
\Biggr)
=\biggl(\frac{2\Lambda^{D-4} }{\Gamma_{ K_0} }\biggr)^2
\Biggl(
{
\partial^\mu \Bigl[-
\partial^\sigma \Gamma_{J_a^\mu J_b^\sigma} 
-\frac{v^2}{2}m^2 \Gamma_{J_a^\mu N_{0b}}\Bigr]
}
\nonumber\\&& 
- \frac{v^2}{2}m^2 \Bigl[
{
- \partial^\mu \Gamma_{ N_{0a}J_b^\mu} 
- \frac{v^2}{2}m^2 \Gamma_{N_{0a} N_{0b} }
{-\frac{1}{2} \Gamma_{ N_{ab}}}+ \frac{1}{2} \delta_{ab}\Gamma_{ N_{00}}
}
\Bigr]
\Biggr).
\label{h2.3}
\end{eqnarray}
In the massless case all the terms present in eq. (\ref{h2.3}) are zero.
For nonzero mass the expected contributions should come from
the tadpole integral 
\begin{eqnarray}&&
B_0 \equiv  \frac{1}{ v^2} \int \frac{d^D q}{(2\pi)^D} 
\frac{1}{(q^2-m^2)};
\nonumber\\&& 
\frac{1}{\Lambda^{D-4}  v^2} \int \frac{d^D q}{(2\pi)^D} 
\frac{1}{(q^2-m^2)}
=
- \frac{i}{v^2} \frac{1}{(4\pi)^2}\Gamma(1-\frac{D}{2})m^2
\Bigl(
\frac{m}{\Lambda\sqrt{4\pi}}
\Bigr)^{D-4}
\nonumber\\&& 
=  i \frac{m^2}{v^2} \frac{1}{(4\pi)^2} \Bigl[-\frac{2}{D-4}
+1 - \gamma + \ln(\frac{m^2}{\Lambda^2(4\pi)})
\Bigr] + {\cal O}(D-4) .
\label{h2.4.1}
\end{eqnarray}
In fact one has
\begin{eqnarray}&&
\Gamma^{(1)}_{K_0} = -\frac{3}{2} i B_0
\label{h2.4.2.1}
\\&&
\Gamma^{(1)}_{N_{00}} = 2 \Gamma^{(1)}_{K_0}= -3 i B_0
\label{h2.4.2.2}
\\&& 
p^\mu
\Gamma^{(1)}_{J^\mu_a  J^\nu_{b}} =
-i \frac{\delta_{ab}}{4}p^\mu\int \frac{d^D q}{(2\pi)^D}
\frac{(2q+p)_\mu(2q+p)_\nu}{[(p+q)^2-m^2][q^2-m^2]}
=
-i \frac{v^2 \delta_{ab}}{2} p_\nu B_0
\label{h2.4.2.3}
\\&& 
\Gamma^{(1)}_{ N_{ab}} = i  \delta_{ab}  B_0
\label{h2.4.2.4}
\\&& 
\Gamma^{(1)}_{N_{0a} N_{0b} }= \Gamma^{(1)}_{J_a^\mu N_{0b}}
=\Gamma^{(1)}_{ N_{0b}}=0
\label{h2.4.2.6}
\\&& 
\Gamma^{(1)}_{ J_{a\mu}(p)\phi_{b}  } =  \frac{v^2}{4} B_0p^\mu \delta_{ab}
\label{h2.4.2.7}
\\&& 
\Gamma^{(1)}_{\phi_{a} N_{0b} } =  \delta_{ab}\frac{5}{3} \Gamma^{(1)}_{K_0}
= -  \frac{5}{2}i B_0\delta_{ab}
\label{h2.4.2.8}
\\&& 
\Gamma^{(1)}_{\phi_a \phi_b} = i \delta_{ab}v^2(p^2+m^2 )  B_0.
\label{h2.4.2.5}
\end{eqnarray}
%

\subsection{Two-point Feynman Rules}
In the zero loop approximation the boundary conditions (\ref{alg.2.1}) and 
eqs. (\ref{h2.1}) and  (\ref{h2.2}) provide the Feynman rules
\begin{eqnarray}&&
\Gamma^{(0)}_{J_a^\mu \phi_b} = 
ip_\mu \delta_{ab}\frac{\Lambda^{D-4} v^2}{2}
\nonumber\\&& 
\Gamma^{(0)}_{N_{0a} \phi_b} = 
\delta_{ab}\Lambda^{D-4} 
\nonumber\\&& 
\Gamma^{(0)}_{\phi_a \phi_b} = 
 \delta_{ab}\Lambda^{D-4}v^2(p^2-m^2). 
\label{h2.4}
\end{eqnarray}
%
\section{On the One-loop Counterterms}
\label{sec:one}
The results of the previous sections allow to extract finite quantities
from dimensional regularized amplitudes. Moreover local counterterms
can be constructed so that one has a recursive process of subtraction.
Perturbative unitarity is guaranteed by construction, under the
form of cutting equation (see for instance Ref. \cite{'tHooft:1973pz}).
Moreover the LFE (\ref{alg.3}) and the consistency condition (\ref{alg.8.1})
allows a order-by-order check of the counterterms.
\par
It is of some interest to look for the local solutions of the LFE (\ref{alg.3})
and of its associated homogeneous equation. In fact, when the solutions
are known, one can obtain all the counterterms for the descendant
amplitudes counterterms. Further on we provide an example for the one-loop
approximation. On one side this subject is very instructive, but on the other
side, as it will be clear later on, 
the study of the local invariant solutions becomes very complex
at higher order in the perturbative expansion.
\par
We list some of the  monomials that can be associated to one-loop divergent
amplitudes. 
According to the eq. (\ref{h.2}) they are expected to have 
dimension  4 or less, if we consider the subtraction procedure at $D=4$.
Any monomial that does not contain the $\phi$-field (also implicitly as in  
$F_{a\mu}$ or in $K_A$) is a sterile term, since it cannot be the ancestor 
of any descendant amplitude. These sterile terms cannot be neglected:
some of them take care of the counterterms associated to tadpoles. 
Here are few examples of dimension four constructed by using the bleached
operators (i.e. invariant under the local transformations generated
by ${\cal S}_a (x)$). Moreover global $SU(2)_R$ invariance is imposed. 
The antisymmetric part of $N_{AB}$ does
not appear in the counterterms.
\par
Let us list first the possible counterterms  present in the massless case
\cite{Ferrari:2005va}
{
\begin{eqnarray}
&& {\cal I}_1 = \int d^Dx \, \Big [ D_\mu[F] ( J- F )_\nu \Big ]_a \Big [ D^\mu[F] ( J- F )^\nu \Big ]_a  \, , 
\nonumber \\
&& {\cal I}_2 = \int d^Dx \, \Big [ D_\mu[F] ( J- F )^\mu \Big ]_a \Big [ D_\nu[F] ( J- F )^\nu \Big ]_a  \, , 
\nonumber \\
&& {\cal I}_3 = \int d^Dx \, \epsilon_{abc} \Big [ D_\mu[F] ( J- F )_\nu \Big ]_a \Big ( J^\mu_b - F^\mu_b \Big ) \Big ( J^\nu_c -F^\nu_c \Big ) \, ,  \nonumber \\
&& {\cal I}_4 = 
\int d^Dx \, \Big (K_A\phi_A\Big )^2
\nonumber \\&&
~~~~~=\int d^Dx \, \Big ( K_0\phi_0 
-\frac{\phi_a}{\Lambda^{D-4} }\TG^{(0)}_{\phi_a}\Big )^2 \, , 
\nonumber \\
&& {\cal I}_5 = \int d^Dx \, \Big ( K_A\phi_A\Big ) \Big (J^\mu_c -F^\mu_c\Big )^2
\nonumber \\
&&
~~~~~=\int d^Dx \, \Big (  K_0\phi_0  -\frac{\phi_a}{\Lambda^{D-4} }\TG^{(0)}_{\phi_a} \Big )
 \Big (J^\mu_c -F^\mu_c\Big )^2 \, , 
\nonumber \\
&& {\cal I}_6 = \int d^Dx \, \Big ( J^\mu_a -F^\mu_a\Big  )^2
 \Big ( J^\nu_b -F^\nu_b \Big )^2 \, , \nonumber \\
&& {\cal I}_7 = \int d^Dx \, \Big ( J^\mu_a -F^\mu_a\Big  )
   \Big ( J^\nu_a -F^\nu_a\Big  ) 
   \Big ( J_{b\mu} -F_{b\mu} \Big  )
   \Big ( J_{b\nu} -F_{b\nu} \Big  )
\, .
\label{alg.7}
\end{eqnarray}
}
Notice the identity
\begin{eqnarray}&&
2({\cal I}_1 -{\cal I}_2 )-4{\cal I}_3+({\cal I}_6-{\cal I}_7 )
= \int d^D x {\cal G}_{a\mu\nu}[{\mathfrak J}]{\cal G}_a^{\mu\nu}[{\mathfrak J}]
= \int d^D x {\cal G}_{a\mu\nu}[J]{\cal G}_a^{\mu\nu}[J],
\label{alg.8}
\end{eqnarray}
where
\begin{eqnarray}
{\cal G}_{a\mu\nu}[J] \equiv \partial_\mu J_{a\nu}-\partial_\nu J_{a\mu}
+\epsilon_{abc}J_{b\mu}J_{c\nu}.
\label{alg.8.1.2}
\end{eqnarray}
The right hand term in eq. (\ref{alg.8}) is sterile: no descendant terms are generated.
The calculation \cite{Ferrari:2005va} for the massless nonlinear sigma model gives 
\begin{eqnarray}&&
\widehat \G^{(1)}\Bigr|_{\rm MASSLESS} 
= \frac{1}{D-4} \frac{\Lambda^{D-4}}{(4\pi)^2}\Big [
- \frac{1}{12}   \Big (
{\cal I}_1 - {\cal I}_2 -  {\cal I}_3 \Big ) 
+  \frac{1}{48} 
\Big ( {\cal I}_6 + 2 {\cal I}_7 \Big ) 
\nonumber \\& &
+ 
 \frac{3}{2} \frac{1}{v^4 } {\cal I}_4
+  \frac{1}{2} \frac{1}{v^2} {\cal I}_5 \Big ]. 
\label{alg.7.1}
\end{eqnarray}
These counterterms are expected to be present in the same combination in the massive case,
since no tadpoles contribute to the invariants ${\cal I}_1-{\cal I}_7$.
\par
For later use we display the local invariant
\begin{eqnarray}&&
K_A\phi_A \Bigr|_{N_{00} \to N_{00}-\frac{m^2v^2}{2}}
=\frac{ K_0}{\phi_0} - \phi_a \biggl(-v^2\Box\phi_a 
\nonumber \\& &
- v^2 \int d^D x \Bigl[\frac{J_b^\mu}{4}\frac{\delta F_{b\mu}}{\delta \phi_a}\Bigr]
+N_{aB}\phi_B + \phi_BN_{Ba}-2\phi_a (N_{00}{+}\frac{m^2v^2}{2})
\biggr)
\label{alg.7.2}
\end{eqnarray}
and its ancestor content
\begin{eqnarray}
K_A\phi_A\Bigr|_{\vec\phi=0} = K_0 .
\label{alg.7.3}
\end{eqnarray}
\par
The massive case requires the introduction of the sources $N_{AB}$ thus we get
a numerous set of new invariants. At the one-loop level we have the
 local invariants
candidates for $\widehat\TG^{(1)}$
\begin{eqnarray}
&&
{\cal I}_{8} = \int d^Dx \,\,\,N_{CC}
   \nonumber \\&&
{\cal I}_{9} = \int d^Dx \,\,\,K_A \phi_A
   \nonumber \\&&
{\cal I}_{10} = \int d^Dx \,\,\,\Big (J^\mu_c -F^\mu_c\Big )^2
   \nonumber \\&&
{\cal I}_{11} = \int d^Dx \,\,\, \phi_A N_{AB} \phi_B 
   \nonumber \\&&
{\cal I}_{12} = \int d^Dx \, \,\, \phi_A \Big (N_{AB}+N_{BA}\Big  )\Big (N_{BC}+N_{CB}\Big  ) \phi_C 
   \nonumber \\&&
{\cal I}_{13} = \int d^Dx \,  \,\, \phi_A N_{AB}\phi_B \phi_C N_{CD}\phi_D
   \nonumber \\&&
{\cal I}_{14} = \int d^Dx \, \,\, K_C \phi_C  \phi_A N_{AB}\phi_B
   \nonumber \\&&
{\cal I}_{15} = \int d^Dx \,\,\, K_A  \Big (N_{AB}+N_{BA}\Big  )  \phi_B ,
   \nonumber \\&& \,\,\,\,
{K_A  \Big (N_{AB}+N_{BA}\Big  )\Bigr|_{\vec\phi=0}=2K_0N_{00}
-(\frac{v^2}{2}\partial_\mu J_a^\mu + N_{0a} + N_{a0})(N_{0a} + N_{a0})
}
   \nonumber \\&&
{\cal I}_{16} = \int d^Dx \,\,\,  \phi_A N_{AB} \phi_B\Big (J^\mu_c -F^\mu_c\Big )^2
   \nonumber \\&&
{\cal I}_{17} = \int d^Dx \, \,\, K_C \phi_C   N_{AA}
   \nonumber \\&&
{\cal I}_{18} = \int d^Dx \, \,\,   \Big ( N_{AB} N_{AB}+ N_{AB} N_{BA}\Big )
   \nonumber \\&&
{\cal I}_{19} = \int d^Dx \, \,\, N_{AA} \Big (J^\mu_c -F^\mu_c\Big )^2
   \nonumber \\&&
{\cal I}_{20} = \int d^Dx \, \,\, (N_{AB}+N_{BA}) \, Tr\, \Bigl\{\Omega ^\dagger (F^\mu-J^\mu)\Omega 
\tau^\dagger_A (F_\mu-J_\mu)\tau_B \Bigr\}
   \nonumber \\&&
{{\cal I}_{21} = \int d^Dx \, \,\, K_A K_A,\qquad
K_A K_A\Bigr|_{\vec\phi=0}=K_0^2
+(\frac{v^2}{2}\partial_\mu J_a^\mu + N_{0a} + N_{a0})^2
}
\, .
\label{alg.9}
\end{eqnarray}
Let us find the ancestor variables content of the invariant ${\cal I}_{20}$ by using 
\begin{eqnarray}
[\tau_A^\dagger,\frac{\tau_b}{2} ]= \delta_{Ax}\epsilon_{xbc}\tau_c,\qquad (N_{AB}+N_{BA})\tau_A^\dagger\tau_B
= 2 N_{AB}.
\label{alg.9.2}
\end{eqnarray}
We have
\begin{eqnarray}&&
{\cal I}_{20}\Bigr|_{\vec\phi=0} =(N_{AB}+N_{BA}) \, Tr\, \Bigl\{ J^\mu
\tau^\dagger_A J_\mu\tau_B \Bigr\}
\nonumber \\&&
= (N_{AB}+N_{BA}) \biggl(\frac{1}{2}
J_a^\mu J_{a\mu} \delta_{AB}
+
\delta_{Ax}\epsilon_{xyz}J_{y\mu} \, Tr\, \Bigl\{J^\mu \tau_z
\tau_B \Bigr\}
\biggr)
\nonumber \\&&
= (N_{AB}+N_{BA}) \biggl(\frac{1}{2}
J_a^\mu J_{a\mu} \delta_{AB}
-
\delta_{Ax}\epsilon_{xyz}J_{y\mu} \epsilon_{y' z x'}\delta_{x'B}\, J_{y'}^\mu 
\biggr)
\nonumber \\&&
= (N_{AB}+N_{BA}) \biggl(\frac{1}{2}
J_a^\mu J_{a\mu} \delta_{AB}
+
\delta_{Ax}J_{y\mu}(\delta_{xy'}\delta_{yx'}-\delta_{xx'}\delta_{yy'}) \delta_{x'B}\, J_{y'}^\mu 
\biggr)
\nonumber \\&&
= 
N_{AA}J_a^\mu J_{a\mu} + 2 N_{ab} J_{a\mu} J_{b}^\mu - 2 N_{bb}  J_a^\mu J_{a\mu}
\nonumber \\&&
= 
(N_{00}-N_{bb} )J_a^\mu J_{a\mu} + 2 N_{ab} J_{a\mu} J_{b}^\mu .
\label{alg.9.3}
\end{eqnarray}
\par
In Appendix \ref{app:eva} we evaluate the coefficients of the 
invariants listed in eq. (\ref{alg.9}).
Finally the tadpoles, originating from the mass term,
necessitate the following counterterms at one-loop, that add to those
for the massless nonlinear sigma model in eq. (\ref{alg.7.1})
\begin{eqnarray}
\widehat \G^{(1)}\Bigr|_{\rm TADPOLES} 
&= &\frac{1}{D-4} \frac{\Lambda^{D-4}}{(4\pi)^2}\Big [
-2\frac{m^2}{v^2}   {\cal I}_{8}
+2 \frac{m^2}{v^2}{\cal I}_{11}-\frac{1}{v^4}   {\cal I}_{12}
+8 \frac{1}{v^4} {\cal I}_{13}
\nonumber \\& & 
+8 \frac{1}{v^4}  {\cal I}_{14}
+\frac{2}{v^2} {\cal I}_{16}-2 \frac{1}{v^4}   {\cal I}_{17}
+\frac{1}{v^4}   {\cal I}_{18}-\frac{1}{2v^2} {\cal I}_{19}
-\frac{1}{2v^2} {\cal I}_{20}\Big ].
\label{alg.10.2.18}
\end{eqnarray}
From the expression in eqs. (\ref{alg.7.1}) and (\ref{alg.10.2.18}) one can get
all the one-loop counterterms by taking the relevant functional
derivatives. As an example one gets
\begin{eqnarray}&&
\widehat \G^{(1)}_{\phi_a\phi_b}
= \delta_{ab}\frac{1}{D-4} \frac{\Lambda^{D-4}}{(4\pi)^2}\Big [
-2  m^4
+ 2  m^4
-8 m^4 
-8  m^2(p^2 {-m^2})
\nonumber \\& & 
+8 m^2  p^2
+2  m^2(p^2 {-m^2})
-2m^2  p^2
-2m^2  p^2\Big ]
\nonumber \\& & 
= \delta_{ab}\frac{1}{D-4} \frac{\Lambda^{D-4}}{(4\pi)^2}\Big [
- 2  m^2 (p^2 {+m^2})\Big ].
\label{alg.10.2.19}
\end{eqnarray}
which agrees with the direct calculation in eq. (\ref{h2.4.2.5}).
%

\section{Conclusions}
\label{sec:conc}
The subtraction strategy, recently developed for
the nonlinear sigma model and for the nonabelian gauge theories,
is implemented here for the {\sl massive} nonlinear sigma model 
(without fermions in this work).
In the present paper the technique has been applied to the
simple case of $SU(2)\otimes SU(2)$. The extension to other groups
of transformations
is expected to be straightforward. 
\par
The main tool is the LFE for the
effective action, derived from the invariance properties of the 
path integral measure. The presence of a mass term requires
the introduction of more  sources coupled to additional composite
operators. However this fact brings only to a more complex algebra,
without diminishing the power of the LFE. The hierarchy still
works so that all 1PI-amplitudes with external  field-parameters
$\vec\phi$ can be derived from those with only composite operators.
The hierarchy allows to organize, at every order of the loop
expansion, the infinite set of divergent amplitudes, so that their
divergent parts can be expressed in term of a finite number
of divergent amplitudes. 
The subtraction algorithm exploits
this powerful property and it is based on the dimensional regularization
and on the subtraction of the {\sl sole} pole parts (no finite adjustments
are allowed). The WPC is shown to be stable under the subtraction
procedure. The linearized  LFE suggests the use of powerful
{\sl local} $SU(2)$ symmetry properties (gauge-type) in order to study the
form of the counterterms. The bleaching technique is very
useful since it maps all external source monomials into
invariant quantities, hence very handy objects for the final aim:
the construction of the counterterms.
\par 
The resulting perturbative expansion yields amplitudes
that depend on the mass $m$, on the Spontaneous Symmetry Breaking
parameter $v$ and on the scale of the radiative corrections $\Lambda$.


\section*{Acknowledgments} 
The author is pleased to thank the Center for Theoretical Physics at MIT,
Massachusetts, where he had the possibility to work on this research.

\appendix

\section{More on Hierarchical Relations}
\label{app:A}
We provide more relations and Feynman rules for the three- and four-point
amplitudes. The approach is outlined in Section \ref{sec:h2}.

\subsection{For the Three-point Functions}
We perform further derivatives of  eq. (\ref{lfe.5}) and  we get the following relations
among the ancestor amplitudes
\begin{eqnarray}&&
\partial^\mu \Gamma_{J_a^\mu J_b^\nu J_c^\sigma} + \epsilon_{abc'} \Gamma_{J_{c'}^\nu J_c^\sigma} 
+ \epsilon_{acc'} \Gamma_{J_{c'}^\sigma J_b^\nu} 
- \frac{1}{2}m^2 \Gamma_{N_{0a}J_b^\nu J_c^\sigma}
\nonumber\\&&
-\frac{1}{2\Lambda^{D-4} }  \Gamma_{ K_0} {\Gamma_{\phi_a J_b^\nu J_c^\sigma}}
 \,   =0.
\label{h3.1}
\end{eqnarray}
\begin{eqnarray}&&
\partial^\mu \Gamma_{J_a^\mu J_b^\nu N_{0c}} + \epsilon_{abc'} \Gamma_{J_{c'}^\nu N_{0c}} 
- \frac{1}{2}m^2 \Gamma_{N_{0a}J_b^\nu N_{0c}}
- \frac{1}{2}  \Gamma_{N_{ca} J_b^\nu} + \frac{1}{2}  \delta_{ac}\Gamma_{ N_{00}J_b^\nu}
\nonumber\\&&
+\frac{1}{2}\epsilon_{acc'} \Gamma_{ N_{0c'}J_b^\nu}
-\frac{1}{2\Lambda^{D-4} v^2}  \Gamma_{ K_0} {\Gamma_{\phi_a J_b^\nu N_{0c}}} \,   =0.
\label{h3.2}
\end{eqnarray}
\begin{eqnarray}&&
\partial^\mu \Gamma_{J_a^\mu N_{0b}N_{0c }} 
- \frac{1}{2}m^2 \Gamma_{N_{0a} N_{0b}N_{0c }}
- \frac{1}{2}  \Gamma_{N_{ba}N_{0c }} 
- \frac{1}{2}  \Gamma_{N_{ca}N_{0b}} 
+ \frac{1}{2} \delta_{ab} \Gamma_{ N_{00}N_{0c }}
+ \frac{1}{2}  \delta_{ac} \Gamma_{ N_{00}N_{0b }}
\nonumber\\&&
+\frac{1}{2}\epsilon_{abc'} \Gamma_{ N_{0c'}N_{0c }}
+\frac{1}{2}\epsilon_{acc'} \Gamma_{ N_{Bc'}N_{0b }}
-\frac{1}{2\Lambda^{D-4} v^2}  \Gamma_{ K_0} {\Gamma_{\phi_a N_{0b}N_{0c }}}
 \,   =0.
\label{h3.2.1}
\end{eqnarray}
If one $\phi$ - derivative is taken, one has
\begin{eqnarray}&&
{\partial^\mu \Gamma_{J_a^\mu J_b^\nu \phi_c}} + \epsilon_{abc'} \Gamma_{J_{c'}^\nu\phi_c} 
- \frac{1}{2}m^2 {\Gamma_{N_{0a}J_b^\nu \phi_c}}
\nonumber\\&&
-\frac{1}{2\Lambda^{D-4} v^2}  \Gamma_{ K_0} {\Gamma_{\phi_a J_b^\nu \phi_c}}+ 
\frac{1}{2}\epsilon_{acc'} \Gamma_{\phi_{c'} J_b^\nu}
 \,   =0.
\label{h3.3}
\end{eqnarray}
\begin{eqnarray}&&
{\partial^\mu \Gamma_{J_a^\mu N_{0b }\phi_c }} 
- \frac{1}{2}m^2 {\Gamma_{N_{0a}N_{0b }\phi_c}}
- \frac{1}{2}  \Gamma_{N_{ba}\phi_c } 
\nonumber\\&&
+\frac{1}{2}\epsilon_{abc'} \Gamma_{ N_{0c'}\phi_c}
-\frac{1}{2\Lambda^{D-4} v^2}  \Gamma_{ K_0} {\Gamma_{\phi_a N_{0b }\phi_c }}+ 
\frac{1}{2}\epsilon_{acc'} \Gamma_{\phi_{c'}N_{0b }}
 \,   =0.
\label{h3.4}
\end{eqnarray}
\begin{eqnarray}&&
{\partial^\mu \Gamma_{J_{a_1}^\mu N_{a_2a_3 }\phi_{a_4} }}
- \frac{1}{2}m^2 {\Gamma_{N_{0a_1}N_{a_2a_3 }\phi_{a_4}}}
+ \frac{1}{2}  \delta_{a_1a_2} \Gamma_{N_{0a_3}\phi_{a_4} } 
+ \frac{1}{2}  \delta_{a_1a_3} {\Gamma_{ N_{a_2 0}\phi_{a_4}}}
\nonumber\\&&
-\frac{1}{2\Lambda^{D-4} v^2}  \Gamma_{ K_0} 
{\Gamma_{\phi_{a_1} N_{a_2a_3 }\phi_{a_4}}}+ 
\frac{1}{2}\epsilon_{{a_1}{a_4}c'} \Gamma_{\phi_{c'}N_{a_2a_3 }}
 \,   =0.
\label{h3.4.1}
\end{eqnarray}
Two $\phi$ - derivatives yields
\begin{eqnarray}&&
{\partial^\mu \Gamma_{J_a^\mu \phi_b \phi_c} }
- \frac{1}{2}m^2 {\Gamma_{N_{0a }\phi_b \phi_c}}
-\frac{1}{2\Lambda^{D-4} v^2}  \biggl(\Gamma_{ K_0} \Gamma_{{\phi_a \phi_b \phi_c}}
+\Gamma_{ K_0 {\phi_b}} \Gamma_{{\phi_a \phi_c}}
\nonumber\\&&
+\Gamma_{ K_0 {\phi_c}} \Gamma_{{\phi_a \phi_b}}
+\Gamma_{ K_0 {\phi_c\phi_b} } \Gamma_{{\phi_a }}
\biggr)+ 
\frac{1}{2}\epsilon_{abc'} \Gamma_{\phi_{c'}\phi_c}+ 
\frac{1}{2}\epsilon_{ac c'} \Gamma_{\phi_{c'}\phi_b}
 \,   =0.
\label{h3.5}
\end{eqnarray}
Thus one can obtain all the amplitudes involving the $\phi$ - fields
($\Gamma_{{\phi_a \phi_b \phi_c}}=0$).
%
\subsection{Three-point Feynman Rules}
From eq. (\ref{lfe.5})  we get
\begin{eqnarray}
\Gamma^{(0)}_{K_0\phi_a \phi_b} = 
-  \delta_{ab}\Lambda^{D-4}
\label{h3.6}
\end{eqnarray}
From eqs. (\ref{h3.3})  and   (\ref{h3.4}) we get
\begin{eqnarray}&&
\Gamma^{(0)}_{J^\mu_a\phi_b \phi_c} 
= - \frac{i}{2}\Lambda^{D-4}v^2\epsilon_{abc}(p_b-p_c)_\mu
\nonumber\\&&
\Gamma^{(0)}_{ N_{0a }\phi_b \phi_c}=0
\nonumber\\&&
\Gamma^{(0)}_{\phi_{a_1} N_{a_2a_3 }\phi_{a_4}}
= \Lambda^{D-4}(\delta_{a_1a_2}  \delta_{a_4a_3}+ \delta_{a_1a_3}\delta_{a_4a_2})
\label{h3.7}
\end{eqnarray}
%
\subsection{For the four-point Functions}
We consider also the four-point functions, but we derive only
the relations that are necessary in order to get the Feynman rules used in the present paper.
We get the following  identities from eq. (\ref{lfe.5})
\begin{eqnarray}&&
{
-\partial^\mu \Gamma_{J_{a_1}^{\mu_1}J_{a_2}^{\mu_2}\phi_{a_3}\phi_{ a_4}} }
- \epsilon_{a_1a_2c} \Gamma_{J_c^{\mu_2}\phi_{a_3}\phi_{ a_4}} 
{
+ \frac{v^2}{2} m^2 \Gamma_{N_{0a_1}J_{a_2}^{\mu_2}\phi_{a_3}\phi_{ a_4}}}
\nonumber\\&&
+\frac{1}{2\Lambda^{D-4} }  \Gamma_{ K_0} \Gamma_{\phi_{a_1}J_{a_2}^{\mu_2}\phi_{a_3}\phi_{ a_4}}
+\frac{1}{2\Lambda^{D-4} }  \Gamma_{K_0 \phi_{a_3}\phi_{ a_4}} \Gamma_{\phi_{a_1}J_{a_2}^{\mu_2}}
{
+\frac{1}{2\Lambda^{D-4} }  \Gamma_{ K_0J_{a_2}^{\mu_2}\phi_{a_3}} \Gamma_{\phi_{a_1}\phi_{ a_4}}}
\nonumber\\&&
{
+\frac{1}{2\Lambda^{D-4} }  \Gamma_{ K_0J_{a_2}^{\mu_2}\phi_{ a_4}} \Gamma_{\phi_{a_1}\phi_{a_3}}}
- 
\frac{1}{2}\epsilon_{a_1a_3c} \Gamma_{\phi_cJ_{a_2}^{\mu_2}\phi_{ a_4}}
- 
\frac{1}{2}\epsilon_{a_1a_4c}\phi_b \Gamma_{\phi_cJ_{a_2}^{\mu_2}\phi_{ a_3}}
 \,   =0.
\label{h4.1}
\end{eqnarray}
We need also
\begin{eqnarray}&&
{
-\partial^\mu \Gamma_{{J_{a_1}^{\mu_1}N_{0 a_2}\phi_{a_3}\phi_{ a_4}}} 
+ \frac{v^2}{2} m^2 \Gamma_{N_{0a_1}N_{0 a_2}\phi_{a_3}\phi_{ a_4}}
}
+\frac{1}{2}\Gamma_{ N_{a_1a_2}\phi_{a_3}\phi_{ a_4}}
-\frac{1}{2} \delta_{a_1a_2} \Gamma_{ N_{00}\phi_{a_3}\phi_{ a_4}}
\nonumber\\&&
+\frac{1}{2\Lambda^{D-4} }  \Gamma_{ K_0} \Gamma_{\phi_{a_1}N_{0 a_2}\phi_{a_3}\phi_{ a_4}}
+\frac{1}{2\Lambda^{D-4} }  \Gamma_{ K_0\phi_{a_3}\phi_{ a_4}} \Gamma_{\phi_{a_1}N_{0 a_2}}
\nonumber\\&&
{
- 
\frac{1}{2}\epsilon_{a_1a_3c} \Gamma_{\phi_cN_{0a_2}\phi_{ a_4}}
- 
\frac{1}{2}\epsilon_{a_1a_4c}\phi_b \Gamma_{\phi_cN_{0a_2}\phi_{ a_3}}
}
 \,   =0.
\label{h4.2}
\end{eqnarray}
Thus finally we can obtain $\Gamma_{\phi_{a_1}\phi_{a_2}\phi_{a_3}\phi_{ a_4}}$
from
\begin{eqnarray}&&
{
-\partial^\mu \Gamma_{J_{a_1}^{\mu_1}\phi_{a_2}\phi_{a_3}\phi_{ a_4}} }
{
+ \frac{v^2}{2} m^2 \Gamma_{N_{0a_1}\phi_{a_2}\phi_{a_3}\phi_{ a_4}}}
+\frac{1}{2\Lambda^{D-4} }  \Gamma_{ K_0} 
\Gamma_{\phi_{a_1}\phi_{a_2}\phi_{a_3}\phi_{ a_4}}
\nonumber\\&&
+\frac{1}{2\Lambda^{D-4} }\sum_{j=2}^4  \Gamma_{K_0 \phi_{a_{j+1}}\phi_{ a_{j+2}}} \Gamma_{\phi_{a_1}\phi_{a_j}}
 \,   =0.
\label{h4.3}
\end{eqnarray}
%

\subsection{Four-point Feynman Rules}
From the above relations (\ref{h4.1}-\ref{h4.3}) at zero loop we get the
Feynman rules
\begin{eqnarray}&&
 \Gamma^{(0)}_{\phi_{a_1}J_{a_2}^{\mu_2}\phi_{a_3}\phi_{ a_4}}=
\nonumber\\&&
\!\!
i \frac{\Lambda^{D-4} v^2}{2}\biggl(\delta_{a_1a_2}\delta_{a_3a_4}(2p_1+p_2)
+\delta_{a_1a_3}\delta_{a_4a_2}(p_2+2p_4)
+ \delta_{a_1a_4}\delta_{a_2a_3}(2p_3+p_2)
\biggr)_{\mu_2}
.
\label{h4.4}
\end{eqnarray}
and
\begin{eqnarray}
\Gamma^{(0)}_{\phi_{a_1}N_{0 a_2}\phi_{a_3}\phi_{ a_4}}
= - 
\Lambda^{D-4}(\delta_{a_1a_2} \delta_{a_3a_4}+\delta_{a_1a_3}\delta_{a_4a_2}+\delta_{a_1a_4}\delta_{a_2a_3})
\label{h4.5}
\end{eqnarray}
Finally from eqs. (\ref{h4.3}), (\ref{h4.4}) and (\ref{h4.5})
\begin{eqnarray}&&
\Gamma^{(0)}_{\phi_{a_1}\phi_{a_2}\phi_{a_3}\phi_{ a_4}}
=
\nonumber\\&&
\Lambda^{D-4} v^2\biggl(\delta_{a_1a_2}\delta_{a_3a_4}(p_2+p_1)^2
+\delta_{a_2a_3}\delta_{a_4a_1}(p_1+p_4)^2
+ \delta_{a_2a_4}\delta_{a_1a_3}(p_3+p_1)^2
\biggr).
\label{h4.7}
\end{eqnarray}
In this Appendix we have verified that the Feynman rules are those
given by the zero loop effective action $\Gamma^{(0)}$.

\section{Evaluation of the Counterterms}
\label{app:eva}
In this Appendix we evaluate the coefficients of the invariants
listed in eq. (\ref{alg.9}) by comparing their external sources content
(i.e. ancestor amplitudes) with the one-loop calculations
given in eqs. (\ref{h2.4.2.1}-\ref{h2.4.2.6}) and with the coefficients
in eq. (\ref{alg.7.1}).
\subsection{Ancestor Invariants}
It should be reminded that the counterterms in eqs. (\ref{alg.7.1}) and  
(\ref{alg.9}) are used as extra Feynman rules after the final substitution
$N_{00}\to N_{00}+\frac{1}{2}m^2v^2$.
\begin{itemize}
\item \underline{$N-J$} amplitudes. 
Since at the one-loop level there is no $\widehat\Gamma^{(1)}_{J^\mu_{b}N_{0a}}$, we have
\begin{eqnarray}
  a_{15}=2 a_{21}.
\label{eva.10.2.11}
\end{eqnarray}
in fact, as shown in eq. (\ref{alg.5}), $K_A$ contains both $N_{0a}$ and $J^\mu_{b}$.
This term can be neglected since it can be written in terms of other invariants
\begin{eqnarray}
\biggl(2 {\cal I}_{15} + {\cal I}_{21} \biggr)\Bigr|_{\vec\phi=0}
= \biggl(\frac{v^4}{4}{\cal I}_{2}+{\cal I}_{4} - {\cal I}_{12}+4{\cal I}_{14} 
\biggr)\Bigr|_{\vec\phi=0}.
\label{eva.10.2.11.1}
\end{eqnarray}
%

\item \underline{$N-J-J$ amplitudes}. We consider the generic combination 
\begin{eqnarray}
a_{10} {\cal I}_{10} + a_{16}{\cal I}_{16} + a_{19}{\cal I}_{19}+ a_{20}{\cal I}_{20}.
\label{eva.10.2}
\end{eqnarray}
At one-loop level we have
\begin{eqnarray}&&
\widehat\Gamma^{(1)}_{N_{00}J_c^\mu J_d^\nu}= 2 \widehat\Gamma^{(1)}_{K_0J_c^\mu J_d^\nu}
=\frac{2}{v^2} g_{\mu\nu}\delta_{cd}
\nonumber\\&&
\widehat\Gamma^{(1)}_{N_{ab}J_c^\mu J_d^\nu}=-(\delta_{ac}\delta_{bd}+\delta_{ad}\delta_{bc})\frac{1}{v^2} g_{\mu\nu}.
\label{eva.12.0}
\end{eqnarray}
Thus we get
\begin{eqnarray}&&
a_{16}+a_{19}+a_{20} = \frac{1}{v^2}
\nonumber\\&&
a_{19}=a_{20}
\nonumber\\&&
a_{20} = - \frac{1}{2v^2}
\nonumber\\&&
a_{16} = \frac{2}{v^2}.
\label{eva.12.0.1}
\end{eqnarray}
%
%
%
\item \underline{$J-J$ amplitudes}. Now we can check the content of eq. (\ref{eva.10.2})
\begin{eqnarray}
a_{10} {\cal I}_{10} + \frac{1}{2v^2}\biggl(4{\cal I}_{16} -{\cal I}_{19}-{\cal I}_{20}\biggr),
\label{eva.10.2.1}
\end{eqnarray}
to be compared with the expected counterterm coming from eq. (\ref{h2.4.2.3}), i.e. $m^2$. We get%
\begin{eqnarray}
2a_{10} {+}  \frac{m^2v^2}{2}\frac{1}{2v^2} 2 \biggl(4 -1-1\biggr)=m^2
\label{eva.10.2.2}
\end{eqnarray}
Thus the final combination is
\begin{eqnarray}
{
\frac{1}{2v^2}\biggl(4{\cal I}_{16} -{\cal I}_{19}-{\cal I}_{20}\biggr).
}
\label{eva.10.2.3}
\end{eqnarray}
%
\item \underline{$N-N$ amplitudes}. 
Now we consider the invariants that contain $N_{AB}$. 
The $N_{00}$ is coupled to $\vec\phi\, ^2$, while $N_{0a}$ to $\vec\phi\, ^2 \phi_a$
and finally $N_{ab}$  to $\phi_a \phi_b$. Thus there is no countertems for $\Gamma_{N_{00}N_{0a}}$
and $\Gamma_{N_{aa}N_{0b}}$. Finally at one loop there no divergence for $\Gamma^{(1)}_{N_{0b}N_{0a}}$.
We group together all invariants with ancestor amplitude containing only $N_{AB}$
\begin{eqnarray}&&
a_{8} {\cal I}_{8}+a_{11}{\cal I}_{11} + a_{12}{\cal I}_{12}+ a_{13}{\cal I}_{13}+ a_{18}{\cal I}_{18}
\label{eva.10.2.4}
\end{eqnarray}
The absence of bilinear terms involving $N_{0a}$ requires the combination
\begin{eqnarray}&&
a_{8} {\cal I}_{8}+a_{11}{\cal I}_{11} + a_{12}({\cal I}_{12}-{\cal I}_{18})+ a_{13}{\cal I}_{13}
\label{eva.10.2.5}
\end{eqnarray}
In fact
\begin{eqnarray}&&
\biggl({\cal I}_{12}-{\cal I}_{18}\biggr)\Big|_{\vec\phi=0}
=\biggl(
2\phi_A N_{AB}N_{BC}\phi_C - N_{AB}N_{BA}
\nonumber\\&&
+\phi_A N_{AB}N_{CB}\phi_C + \phi_A N_{BA}N_{BC}\phi_C  - N_{AB}N_{AB} \biggr)
\Big|_{\vec\phi=0}
\nonumber\\&&
=
2 N_{00}^2 +2N_{0a}N_{a0}-N_{00}^2-2N_{0a}N_{a0}-N_{ab}N_{ba}
+N_{00}^2+N_{0a}N_{0a}
\nonumber\\&&
+N_{00}^2+N_{a0}N_{a0}-N_{00}^2-N_{0a}N_{0a}-N_{a0}N_{a0}-N_{ab}N_{ab}
\nonumber\\&&
= 2 N_{00}^2 -\biggl(N_{ab}N_{ab}+N_{ab}N_{ba}\biggr)
\label{eva.10.2.6}
\end{eqnarray}
Thus we get
\begin{eqnarray}&&
\widehat\Gamma^{(1)}_{ N_{00} N_{00}}=   4 a_{12} + 2 a_{13}=  4 \widehat\Gamma^{(1)}_{K_0K_0}= \frac{12}{v^4}
\nonumber\\&&
\widehat\Gamma^{(1)}_{ N_{aa'}N_{bb'}}= -  2 a_{12}(\delta_{ab}\delta_{a'b'}+\delta_{ab'}\delta_{ba'} )=
 (\delta_{ab}\delta_{a'b'}+\delta_{ab'}\delta_{ba'} )\frac{2}{v^4}
\label{eva.10.2.7}
\end{eqnarray}
i.e.
\begin{eqnarray}
 a_{12} = - \frac{1}{v^4}, \qquad a_{13}= \frac{8}{v^4}.
\label{eva.10.2.8}
\end{eqnarray}
%
\item \underline{$N$ - amplitudes}. 
Finally from eqs. (\ref{h2.4.2.2}) and (\ref{h2.4.2.4}) we get the contribution to $\widehat\Gamma^{(1)}_{ N_{AB}}$
\begin{eqnarray}&&
a_{8} + a_{11} {+} m^2v^2(2 a_{12} +  a_{13}) = 6\frac{m^2}{v^2}
\nonumber\\&&
 a_{8}\delta_{ab} = -2 \frac{m^2}{v^2}\delta_{ab}, \qquad a_{11} =  {2} \frac{m^2}{v^2}.
\label{eva.10.2.9}
\end{eqnarray}
Finally
\begin{eqnarray}&&
- 2 \frac{m^2}{v^2}{\cal I}_{8}+ {2}\frac{m^2}{v^2}{\cal I}_{11} 
- \frac{1}{v^4}({\cal I}_{12}-{\cal I}_{18})+ \frac{8}{v^4}{\cal I}_{13}.
\label{eva.10.2.10}
\end{eqnarray}
%

\item  \underline{$K_0-N$ amplitudes}. We consider the linear combination
\begin{eqnarray}
a_9{\cal I}_{9}+ a_{14} {\cal I}_{14}+ a_{17} {\cal I}_{17}.
\label{eva.10.2.12}
\end{eqnarray}
We impose the conditions using eq. (\ref{alg.7.1})
\begin{eqnarray}&&
\widehat\Gamma^{(1)}_{K_0 N_{00}}= a_{14} + a_{17}= 2 \widehat\Gamma^{(1)}_{K_0K_0}= \frac{6}{v^4}
\nonumber\\&&
\widehat\Gamma^{(1)}_{K_0 N_{ab}}= a_{17}\delta_{ab}= -2\frac{1}{v^4} \delta_{ab}. 
\label{eva.10.2.13}
\end{eqnarray}
We get
\begin{eqnarray}
 a_{17}= - \frac{2}{v^4},\qquad  a_{14}=  \frac{8}{v^4}.
\label{eva.10.2.14}
\end{eqnarray}
The tadpole term in eq. (\ref{h2.4.2.1}) requires
\begin{eqnarray}
 a_{9}{+}\frac{m^2v^2}{2}a_{14}{+}  \frac{m^2v^2}{2} a_{17} =  \frac{3m^2}{v^2},\qquad \Longrightarrow a_{9}= {0}.
\label{eva.10.2.15}
\end{eqnarray}
Finally we have
\begin{eqnarray}
{
\frac{8}{v^4} {\cal I}_{14} - \frac{2}{v^4}{\cal I}_{17}.
}
\label{eva.10.2.16}
\end{eqnarray}
%

\end{itemize}
The result of the Appendix is summarized in eq. (\ref{alg.10.2.18}).



\begin{thebibliography}{99}
\small

\bibitem{Ferrari:2005ii}
  R.~Ferrari,
  JHEP {\bf 0508}, 048 (2005)
  [arXiv:hep-th/0504023].


\bibitem{Weinberg:1978kz}
  S.~Weinberg,
  Physica A {\bf 96} (1979) 327.

\bibitem{Colangelo:1995jm}
  G.~Colangelo,
  Phys.\ Lett.\  B {\bf 350}, 85 (1995)
  [Erratum-ibid.\  B {\bf 361}, 234 (1995)]
  [arXiv:hep-ph/9502285].

\bibitem{Bijnens:1998yu}
  J.~Bijnens, G.~Colangelo and G.~Ecker,
  Phys.\ Lett.\  B {\bf 441}, 437 (1998)
  [arXiv:hep-ph/9808421].


\bibitem{Bijnens:2009zi}
  J.~Bijnens and L.~Carloni,
  Nucl.\ Phys.\  B {\bf 827}, 237 (2010)
  [arXiv:0909.5086 [hep-ph]].




\bibitem{Ferrari:2005va}
  R.~Ferrari and A.~Quadri,
  Int.\ J.\ Theor.\ Phys.\  {\bf 45}, 2497 (2006)
  [arXiv:hep-th/0506220].



\bibitem{Ferrari:2005fc}
  R.~Ferrari and A.~Quadri,
  JHEP {\bf 0601}, 003 (2006)
  [arXiv:hep-th/0511032].

\bibitem{Bettinelli:2006ps}
  D.~Bettinelli, R.~Ferrari and A.~Quadri,
  Int.\ J.\ Theor.\ Phys.\  {\bf 46} (2007) 2560
  [arXiv:hep-th/0611063].
%


\bibitem{Bettinelli:2007zn}
  D.~Bettinelli, R.~Ferrari and A.~Quadri,
  Int.\ J.\ Mod.\ Phys.\  A {\bf 23}, 211 (2008)
  [arXiv:hep-th/0701197].
  
\bibitem{Bettinelli:2007kc}
  D.~Bettinelli, R.~Ferrari and A.~Quadri,
  JHEP {\bf 0703}, 065 (2007)
  [arXiv:hep-th/0701212].





\bibitem{Bettinelli:2007tq}
  D.~Bettinelli, R.~Ferrari and A.~Quadri,
  Phys.\ Rev.\  D {\bf 77}, 045021 (2008)
  [arXiv:0705.2339 [hep-th]].


\bibitem{Bettinelli:2007cy}
  D.~Bettinelli, R.~Ferrari and A.~Quadri,
  Phys.\ Rev.\  D {\bf 77} (2008) 105012
  [arXiv:0709.0644 [hep-th]].





\bibitem{Bettinelli:2008ey}
  D.~Bettinelli, R.~Ferrari and A.~Quadri,
  Int.\ J.\ Mod.\ Phys.\  A {\bf 24} (2009) 2639
  [arXiv:0807.3882 [hep-ph]].


\bibitem{Bettinelli:2008qn}
  D.~Bettinelli, R.~Ferrari and A.~Quadri,
  Acta Phys.\ Pol.\  B {\bf 41}, 597 (2010) 
  [arXiv:0809.1994 [hep-th]].

  

\bibitem{Bettinelli:2009wu}
  D.~Bettinelli, R.~Ferrari and A.~Quadri,
  Phys.\ Rev.\  D {\bf 79}, 125028 (2009)
  [arXiv:0903.0281 [hep-th]].


\bibitem{Ferrari:2009uj}
  R.~Ferrari,
  J.\ Math.\ Phys.\ {\bf 51}, 032305 (2010)
  [arXiv:0907.0426 [hep-th]].
  
\bibitem{'tHooft:1973pz}
  G.~'t Hooft and M.~J.~G.~Veltman,
  NATO Adv.\ Study Inst.\ Ser.\ B Phys.\  {\bf 4} (1974) 177.

  
\end{thebibliography}
\end{document}